\documentclass[prb,twocolumn,amsmath,amssymb]{revtex4}
\usepackage{graphicx}  
\usepackage{natbib}
\usepackage{subfigure}
\usepackage{amsmath}

\begin{document}

\title{Dopant-mediated structural and magnetic properties of TbMnO$_3$}

\author{Vinit Sharma$^{1,2}$, A. McDannald$^{3}$, M. Staruch$^{3}$, R. Ramprasad$^{1,2}$ and M. Jain$^{1,3}$}
\email{menka.jain@uconn.edu}
\affiliation{$^{1}$Institute of Materials Science, University of Connecticut, Storrs, Connecticut 06269, USA\\
$^{2}$Department of Materials Science and Engineering, University of Connecticut, Storrs, Connecticut 06269, USA\\
$^{3}$ Department of Physics, University of Connecticut, Storrs, Connecticut 06269, USA}

\begin{abstract}

Structural and magnetic properties of the doped terbium manganites (Tb,\textit{A})MnO$_3$ (\textit{A} = Gd, Dy and Ho) have been investigated using first-principles calculations and further confirmed by subsequent  experimental studies. Both computational and experimental studies suggest that compared to the parent material, namely,  TbMnO$_3$ (with a magnetic moment of 9.7 $\mu$$_B$ for Tb$^{3+}$) Dy- and Ho- ion substituted TbMnO$_3$ results in an increase in the magnetic moment ($\leqslant{10.6}$$\mu$$_B$  for Dy$^{3+}$ and Ho$^{3+}$). The observed spiral-spin AFM order in TbMnO$_3$ is stable with respect to the dopant substitutions, which modify the Mn-O-Mn bond angles and lead to stronger the ferromagnetic component of the magnetic moment. Given the fact that magnetic ordering in TbMnO$_3$ causes the ferroelectricity, this is an important step in the field of the magnetically driven ferroelectricity in the class of magnetoelectric multiferroics, which traditionally have low magnetic moments due to the predominantly antiferromagnetic order. In addition, the present study reveals important insights on the phenomenological coupling mechanism in detail, which is essential in order to design new materials with enhanced magneto-electric effects at higher temperatures.
\end{abstract}

\maketitle

The phenomenon of inducing electric polarization in a material by the application of an external magnetic field, or vice versa, has attracted considerable attention from the point of view of fundamental science challenges as well as their envisaged appealing potential for applications in the general area of spintronics. \cite{Cheong,Ramesh} Since the discovery of TbMnO$_3$,\cite{TKimura} in which  ferroelectric polarization ($\sim$600$\mu$Ccm$^{-2}$) is induced at its magnetic transition $\sim$27 K and N\`eel  temperature is observed  $\sim$42 K, the orthorhombically distorted perovskite rare-earth manganites, \textit{R}MnO$_3$, have become the most notable class of materials due to their high magnetoelectric coupling (i.e., magnetically induced ferroelectricity).\cite{Zhong} However, due to the relatively low value of electric polarization ($\leqslant{100}$$\mu$Ccm$^{-2}$) and low N\`eel temperature ($\leqslant{27}$K), these materials are not suitable for most technological applications.\cite{Cheong,Ramesh,TKimura}

\begin{figure}
\includegraphics[width=3.3in]{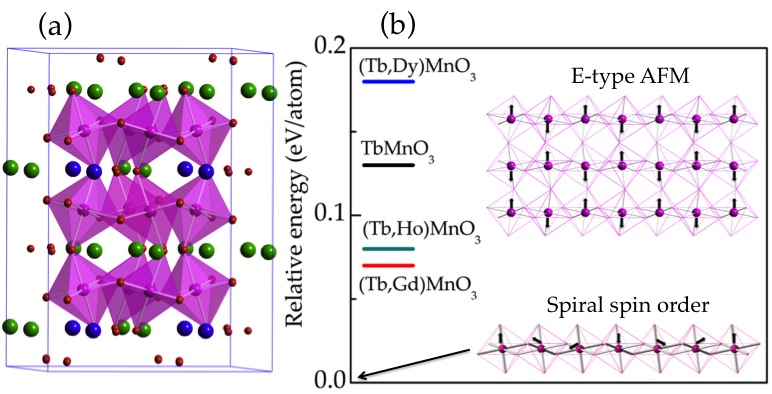}
\caption[Figure1] {(Color online) (\textbf{a}) Crystal structure of perovskite rare-earth manganites TbMnO$_3$ consists of corner sharing MnO$_6$ octahedra, with the Mn ions in the center of the octahedral. The Tb and dopant \textit{A} atoms are at the center of a cube formed by MnO$_6$ octahedra. Green, blue, pink and red spheres represent the Tb, \textit{A}, Mn and O atoms, respectively. (\textbf{b}) DFT based energetics comparing two type of magnetic ordering considered in (Tb,\textit{A})MnO$_3$ (where \textit{ A} = Gd, Dy and Ho) namely E-AFM type and spiral-spin ordering. Considered both E-type AFM spiral-spin ordering are also shown.  For all cases spiral-spin ordering is energetically found favorable, hence as a reference is set to be zero. 
 {\label{fig:Figure1} } }
\end{figure}

Owing to the intrinsic capability of the perovskite structure to host ions of different sizes, chemical doping is considered as a new knob to tune the physical properties of this class of materials.\cite{Sharma} Both computational and experimental techniques  have been used to study a variety of dopants in perovskites \cite{Liu,Ihrig,HIhrig,Chan,Biscaglia,Sambrano,Smyth,Sharma,Nakayama, MBiscaglia,Lewis,Linke,Guo,Mufti,Yang,Perez-Osuna,Cuartero,Drozda,Zhang,Staruch1,Staruch2,Staruch-JPCM,Staruch-APL,Ling,Cuartero2,Cuartero3}. In the particular case of TbMnO$_3$, dopants with different ionic size and different oxidation state such as Ru,\cite{Guo} Ca,\cite{Mufti,Staruch-JPCM} Sr,\cite{Staruch-JPCM}   Na,\cite{Yang} Al,\cite{Perez-Osuna} Sn,\cite{Drozda} La,\cite{Zhang} Dy,\cite{Staruch1} Y,\cite{Staruch1} and Ho\cite{Staruch1,Staruch-APL} have been introduced at the Tb site to tune the magnetic properties. On the other hand, dopants with small ionic size such as Cr,\cite{Staruch2} Sc,\cite{Cuartero,Cuartero2} Co,\cite{Cuartero3} or Fe\cite{Ling} have been introduced at the Mn site to increase  the magnetic transition temperature and to induce ferromagnetic (FM) interactions. 


At ambient conditions, the perovskite  \textit{R}MnO$_3$ crystalizes in either orthorhombic or hexagonal phases.\cite{Staruch1, Staruch2} With  large ionic  size \textit{R} (such as in the cases of Pr, Nd, Tb and  Dy) the orthorhombic phase (\textit{Pbnm}) is found to be stable (Fig 1(a)), while with small ionic size \textit{R} (such as in Ho,  Er,  Yb, and Lu) \textit{R}MnO$_3$ stabilizes in the hexagonal phase (\textit{P63cm}, at room temperature). The orthorhombic \textit{R}MnO$_3$ with small \textit{R} ions (Ho and Lu) can also be synthesized in a metastable orthorhombic phase by high pressure synthesis technique\cite{Wood}  or special chemical methods.\cite{Quezel} Owing to the several magnetic transitions occurring at low temperatures, \textit{R}MnO$_3$, (\textit{R}=Tb, Dy, Lu) with orthorhombic distorted perovskite structure have recently been identified as candidates for magnetic refrigeration utilizing the magnetocaloric effect (MCE).\cite{Zhong} However, due to the relatively broad temperature dependent peak in magnetic entropy change, the under-lying phenomenon governing the origin of such a large magnetocaloric effect (MCE) is still unclear. Since, one can treat the magnetic structure and lattice distortions simultaneously  in first principles computations, these techniques can be used to understand and explain the phenomenological coupling mechanism and can provide necessary insights to design new materials with enhanced MCE operateable at higher temperatures. 

In this contribution, using first-principles density-functional theory (DFT) based methods we have carried out a careful investigation of structural, electronic, and magnetic properties of pure and doped terbium manganites (Tb,\textit{A})MnO$_3$ (\textit{A} = Gd, Dy and Ho). First, we have compared the relative energetics of two types of magnetic ordering anticipated in (Tb,\textit{A})MnO$_3$, namely, the E-AFM type (observed in HoMnO$_3$) and the spiral-spin ordering (observed in TbMnO$_3$). Next, we have studied the effect of dopant \textit{A} (where \textit{ A} = Gd, Dy and Ho) on the electronic and magnetic properties of (Tb,\textit{A})MnO$_3$. Finally, we have compared our computational results with experimentally obtained magnetic properties of pure and doped TbMnO$_3$ bulk ceramics. The present study provides an improved understanding of both the spin structure and magnetization of doped systems and suggests that the substitution of rare-earth ions with R-ions of magnetic  moments  directly affects the ordering of the Tb$^{3+}$ moments and indirectly affects the Mn ordering (causing an increased ferromagnetic component).

The DFT calculations were carried out using the projector augmented plane-wave basis functions as embodied in the Vienna \emph{ab initio} simulation package (VASP).\cite{Kresse-PRB,Kresse-CMS,Kresse-PRB1}  The exchange-correlation interaction was treated within the generalized gradient approximation (GGA) using the Perdew-Burke-Ernzerhoff (PBE) functional.\cite{PBE} It is worth mentioning that for many multiferroic materials such as TbMnO$_{3}$,\cite{Malashevich,Malashevich1} HoMnO$_{3}$,\cite{Picozzi} and CaMn$_7$O$_{12},$\cite{Lu} the first-principles results are very sensitive to the choice of the on-site Coulomb energy \textit{U}.\cite{Malashevich,Malashevich1,Picozzi,Lu,vsharma} Our tests with different \textit{U} values confirm that for TbMnO$_{3}$ \textit{U} = 2 eV for the 3\textit{d} states of Mn reproduces the experimental band gap value of 0.5 eV.\cite{Cui} Hence,  all the  calculations are performed using \textit{U} = 2 eV. \cite{Staruch2}  A Monkhorst-Pack \emph{k}-point mesh\cite{Monkhorst-Pack} of 3$\times$3$\times$2 is employed to produce converged results within 0.1 meV per formula unit. In all calculations spin-orbital coupling (SOC) was included. In doped TbMnO$_3$ studies, one Tb atom was substituted with dopants Gd, Dy, or Ho, thereby resulting in a dopant concentration of 25\%.

\begin{table*}
\tiny
\caption{Summary of the measured structural and magnetic properties of TbMnO$_3$ and Tb$_{0.67}$\textit{A}$_{0.33}$MnO$_3$ (\textit{A} = Gd, Dy and Ho). Quantities computed using  DFT calculations using the GGA+U methods are provided in parenthesis.} 
\begin{tabular}{c|ccc|ccc|ccc|ccc}
  \hline
  \hline
\cline{1-13}
Atomic & \multicolumn{3}{c|}{TbMnO$_3$} & \multicolumn{3}{c|}{Tb$_{0.67}$Gd$_{0.33}$MnO$_3$}  & \multicolumn{3}{c|}{Tb$_{0.67}$Dy$_{0.33}$MnO$_3$} & \multicolumn{3}{c}{Tb$_{0.67}$Ho$_{0.33}$MnO$_3$} \\
\cline{2-13}
 coordinates         &  x         & y         & z      &  x      & y   & z    & x         & y      & z    & x         & y      & z   \\
\hline
Tb/\textit{R} (4\textit{c})   & 0.98 (0.98)  & 0.08 (0.08) & 0.25 (0.25) & 0.98 (0.98)  & 0.08 (0.08) & 0.25 (0.25)  & 0.98 (0.98) & 0.08 (0.08)    & 0.25 (0.25)     &  0.99 (0.98)    &  0.08 (0.08)  & 0.25 (0.25)     \\
O1 (4\textit{c})                 & 0.08 (0.11)  & 0.50 (0.46) & 0.25 (0.25) & 0.08 (0.12)  & 0.50 (0.46) & 0.25 (0.25)  & 0.09 (0.12) & 0.51 (0.46)    & 0.25 (0.25)     &  0.08 (0.12)    &  0.50 (0.46)  & 0.25 (0.25)    \\
O2(4\textit{c})                  & 0.69 (0.70)  & 0.33 (0.32) & 0.05 (0.05) & 0.69 (0.70)  & 0.33 (0.33) & 0.05 (0.05)  & 0.70 (0.70) & 0.33 (0.33)    & 0.05 (0.05)     &  0.70 (0.70)    &  0.32 (0.33)  & 0.05 (0.05)    \\
Mn (4\textit{c})                 & 0.00 (0.00)  & 0.00 (0.00) & 0.50 (0.50) & 0.00 (0.00)  & 0.00 (0.00) & 0.50 (0.50)  & 0.00 (0.00) & 0.00 (0.00)    & 0.50 (0.50)     &  0.00 (0.00)    & 0.00 (0.00)   & 0.50 (0.50) \\
\hline
Structure parameters  & \multicolumn{3}{c|}{}  			     & \multicolumn{3}{c|}{}				    & \multicolumn{3}{c|}{}                      & \multicolumn{3}{c}{}  \\
\textit{a} (\AA) 			& \multicolumn{3}{c|}{5.293 (5.335)} & \multicolumn{3}{c|}{5.310 (5.341)}  & \multicolumn{3}{c|}{5.302 (5.327)} & \multicolumn{3}{c}{5.271 (5.324)} \\
\textit{b} (\AA) 			& \multicolumn{3}{c|}{5.839 (5.884)} & \multicolumn{3}{c|}{5.844 (5.914)}  & \multicolumn{3}{c|}{5.833 (5.895)} & \multicolumn{3}{c}{5.811 (5.892)} \\
\textit{c} (\AA) 			& \multicolumn{3}{c|}{7.402 (7.481)} & \multicolumn{3}{c|}{7.424 (7.471)}  & \multicolumn{3}{c|}{7.418 (7.453)} & \multicolumn{3}{c}{7.378 (7.449)} \\
{}  & \multicolumn{3}{c|}{}  & \multicolumn{3}{c|}{}  & \multicolumn{3}{c|}{} & \multicolumn{3}{c}{} \\
\textit{d}1$_{(Mn-O2)}$ (\AA) & \multicolumn{3}{c|}{2.168 (2.246)} & \multicolumn{3}{c|}{2.168 (2.246)}  & \multicolumn{3}{c|}{2.182 (2.243)} & \multicolumn{3}{c}{2.127 (2.235)} \\
\textit{d}2$_{(Mn-O2)}$ (\AA) & \multicolumn{3}{c|}{1.986 (1.981)} & \multicolumn{3}{c|}{1.986 (1.981)}  & \multicolumn{3}{c|}{1.944 (1.974)} & \multicolumn{3}{c}{1.988(1.975)} \\
\textit{d}$_{(Mn-O1)}$  (\AA) & \multicolumn{3}{c|}{1.907 (1.936)} & \multicolumn{3}{c|}{1.907 (1.936)}  & \multicolumn{3}{c|}{1.920 (1.930)} & \multicolumn{3}{c}{1.896 (1.928)} \\
{}  & \multicolumn{3}{c|}{} & \multicolumn{3}{c|}{}  & \multicolumn{3}{c|}{} & \multicolumn{3}{c}{} \\
Mn - O2 - Mn (deg) & \multicolumn{3}{c|}{144.13 (144.16)} & \multicolumn{3}{c|}{143.8 (144.16)}  & \multicolumn{3}{c|}{145.5 (144.25)} & \multicolumn{3}{c}{144.8 (144.15)} \\
Mn - O1 - Mn (deg) & \multicolumn{3}{c|}{153.5} & \multicolumn{3}{c|}{153.5}  & \multicolumn{3}{c|}{150.0} & \multicolumn{3}{c}{153.3 } \\
{}  & \multicolumn{3}{c|}{} & \multicolumn{3}{c|}{}  & \multicolumn{3}{c|}{} & \multicolumn{3}{c}{} \\
$\chi$$^2$ & \multicolumn{3}{c|}{2.53} & \multicolumn{3}{c|}{2.53}  & \multicolumn{3}{c|}{1.96} & \multicolumn{3}{c}{3.48} \\
{}  & \multicolumn{3}{c|}{} & \multicolumn{3}{c|}{}  & \multicolumn{3}{c|}{} & \multicolumn{3}{c}{} \\
Magnetic Moment ($\mu$$_B$ ) & \multicolumn{3}{c|}{11.31 $\pm$ 0.01(11.29)} & \multicolumn{3}{c|}{9.6$\pm$ 0.01 (9.29) }  & \multicolumn{3}{c|}{11.29$\pm$ 0.01 (11)} & \multicolumn{3}{c}{11.1 $\pm$ 0.01 (10.7)} \\
{}  & \multicolumn{3}{c|}{} & \multicolumn{3}{c|}{}  & \multicolumn{3}{c|}{} & \multicolumn{3}{c}{} \\
\hline
\hline
\end{tabular}
\end{table*}

Powder samples of pure TbMnO$_3$ and doped Tb$_{0.67}$\textit{A}$_{0.33}$MnO$_3$ (\textit{A} = Gd, Dy and Ho) were prepared by a solution based citrate route. High purity nitrate precursors of Tb, Ho, Dy, Gd, and Cr as well as citric acid were dissolved in water and mixed in the stoichiometric ratios. After drying, the powder samples were annealed at 900$^{\circ}$C for 2 hours in an O$_2$ environment. Structural characterization by X-ray diffraction (XRD) was performed using a Bruker D8 diffractometer using Cu- K$\alpha$ radiation. Subsequent Rietveld refinement of the diffraction data was accomplished using the FullProf Suite package.\cite{fullprof} The background was fitted by a linear interpolation between a set of background points with refinable height while refining the structural parameters, and then removed to obtain an accurate goodness of fit. DC magnetization measurements were performed with the Vibrating Sample Magnetometer attachment to the Evercool Physical Property Measurement System from Quantum Design.  

TbMnO$_3$ stabilizes in an orthorhombic perovskite structure (\textit{Pbnm} symmetry) consisting of the corner sharing MnO$_6$ octahedra, with the Mn ions in the center of the octahedral.\cite{Staruch1, Staruch2, Malashevich}  The (Tb,\textit{A}) ions are at the center of a cube formed by MnO$_6$ octahedra as shown in Fig. 1(a). This structure is highly distorted because of the mismatch of (Tb,\textit{A})-O and Mn-O bond distances and the Jahn-Teller (JT) distortion of the Mn-O bonds.  It has been noticed that multiple factors contribute to the spin-orientation of a system. Specially, in orthorhombically distorted  \textit{R}\textit{M}O$_3$ (\textit{R} = rare earth, \textit{M} = transition metal) spin-orientation is dominated by the \textit{M}-ion array and the interaction between the \textit{R} and \textit{M} sublattices, such as antisymmetric exchange interactions,\cite{Su} strong coupling\cite{Yoshii} and anisotropic interactions.\cite{Marshall} Here, as mentioned above two types of the possible magnetic ordering arrangements (below ferroelectric transition temperature) in (Tb,\textit{A})MnO$_3$ are investigated.  Our first-principles calculation based energetics confirm that both pure and  the doped terbium manganites favor the spiral-spin type magnetic ordering, shown in Fig.1 (b).

To get insights on the interplay between orbital- and magnetic- ordering in these manganites we have analyzed  the density of states (DOS). DFT+U+SOC computed total and atom projected DOS of pure and doped terbium manganites are plotted in Fig. 2 (a-d). It is evident from Fig. 2 that the large portion of the DOS (-6 eV to 5 eV) is contributed mainly due to Mn atoms accompanied by Tb atoms. First, we focus on the pure TbMnO$_3$ (Fig. 2a), which is found to be an insulator with energy gap of about 0.5 eV. \cite{Malashevich} The DOS in the conduction band can be explained in terms of optical conductivity spectra and first peak near $\sim$ 1.7 eV is mainly an attribute of the first optical transition as observed in earlier optical conductivity spectra measurements.\cite{Choi} The noteworthy feature of the atom projected DOS is that the peak corresponding to the lowest optical transitions shows a slight shift toward the Fermi level with increasing dopant ionic radius (see Fig. 2 (a-d)).

\begin{figure}
\includegraphics[width=3.0in, height = 2in]{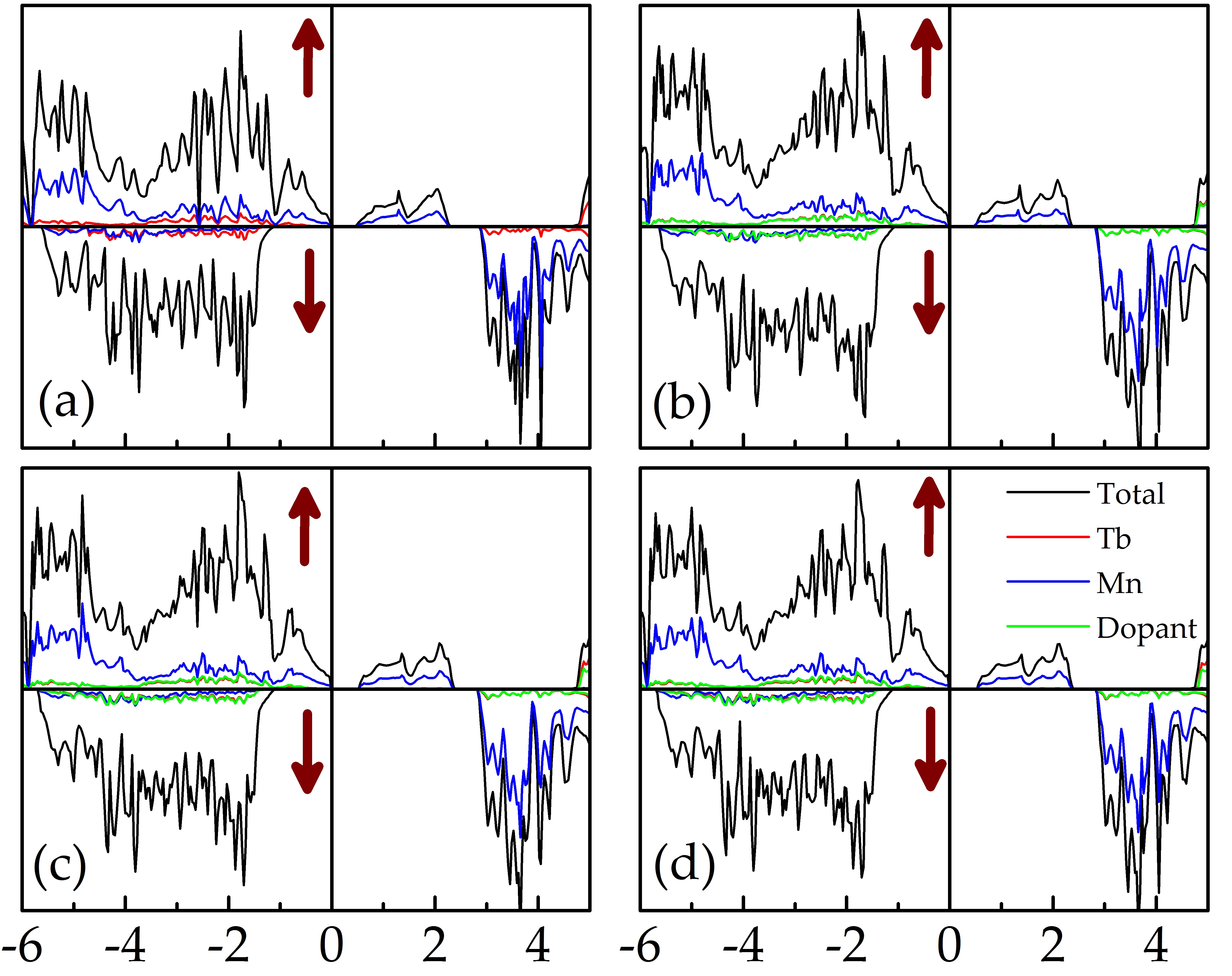}
\caption[Figure1] {(Color online) Calculated total and atom projected density of states (DOS)  of (\textbf{a}) TbMnO$_3$  (\textbf{b}) (Tb,Gd)MnO$_3$ (\textbf{c}) (Tb,Dy)MnO$_3$ (\textbf{d}) (Tb,Ho)MnO$_3$ atoms. In each case, both up and down spin contributions to the DOS are shown.
 {\label{fig:Figure1} } }
\end{figure}

\begin{figure}
\includegraphics[width=3.0in, height = 2in]{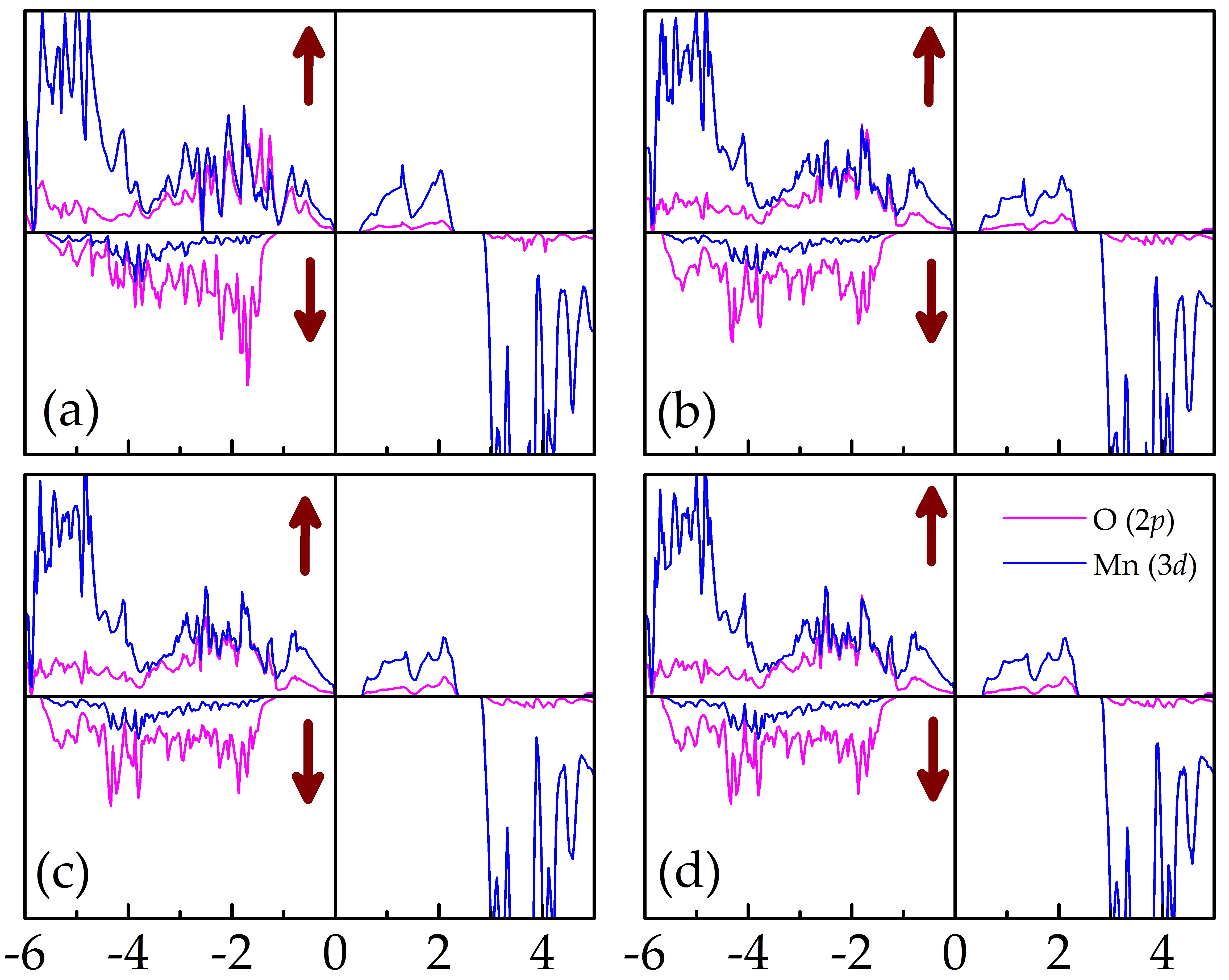}
\caption[Figure1] {(Color online) Calculated orbital projected density of states of O and Mn atoms in (\textbf{a}) TbMnO$_3$ (\textbf{b}) (Tb,Gd)MnO$_3$ (\textbf{c}) (Tb,Dy)MnO$_3$ (\textbf{d}) (Tb,Ho)MnO$_3$ .
 {\label{fig:Figure1} } }
\end{figure}

Next, we examine the partial density of states (PDOS). The orbital-projected DOS in Fig. 3 (a-d) also confirm that in the vicinity of the Fermi-level, the majority of the DOS in the valence band arises from the \textit{d}-states of the Mn atoms, while in the conduction band, the DOS have contributions from both Mn and O atoms. 
The notable point is that the highest occupied level shows O 2\textit{p} character, while the lowest unoccupied level has Mn 3\textit{d} character and mainly dominated by occupation of the \textit{d}$_{(x^{2}-z^{2})}$ orbitals (Fig. 3).
Specifically in the region just above the Fermi level (between 0--2 eV) the DOS arise due to the unoccupied Mn (\textit{d}) $_{3(z^{2}-r^{2})}$ states and above 2 eV the DOS spectrum is dominated by all Mn (3\textit{d}) orbitals. On the other hand, in the valence band below -2 eV joint contribution from O (2\textit{p}) and all Mn (3\textit{d}) orbitals has been observed. It is also evident from Fig. 3 (a-d) that the O (2\textit{p}) states and Mn (3\textit{d}) states further enhance the strong hybridization between the orbital and spin order resulting in the magnetic and structural modulations, consistent with the Jahn-Teller mechanism.\cite{Malashevich}

The XRD $\theta$-2$\theta$ data for the pure TbMnO$_3$ and doped Tb$_{0.67}$\textit{A}$_{0.33}$MnO$_3$ (\textit{A} = Gd, Dy and Ho) samples is shown in Fig. 4(a-d). Peaks corresponding to only the orthorhombic symmetry (space group Pbnm) are observed indicating that all the samples are phase-pure. Rietveld refinement was used to refine the XRD data of all samples. Measured and DFT computed structural parameters (lattice constants, atomic positions, etc.) are listed and compared in Table 1. The lattice parameters for pure TbMnO$_3$ are found to be in agreement with available computational and experimental data.\cite{Staruch1, Staruch2, Malashevich} The slight change in lattice parameters due to the doping can be understood in terms of the ionic size mismatch between host (Tb) and dopants (Gd, Dy and Ho). For instance, the lattice constant was expected to decrease when Ho$^{3+}$ of smaller ionic radius (107.2 pm in 9-coordination) is substituted in the site of Tb$^{3+}$ with larger ionic radius (109.5 pm in 9-coordination). For doped terbium manganites, the average ionic radius of the A-site ion $\textless$ r$_{A}$$\textgreater$ is 108.741 pm and the ionic-size mismatch represented by the variance $\sigma$$^{2}$= $\Sigma$y$_{i}$ $r_{i}^{2}$ - $\textless$r${_A}$$\textgreater$$^{2}$  is 1.17 pm$^{2}$, where r$_{i}$ corresponds to the radii of the various A-site cations and y$_{i}$ to their fractional occupancies. The lowered average radius of the rare-earth ion also explains the reduced value of the Mn-O-Mn bond angle (in-plane) as listed in Table 1, which as previously mentioned has been shown to be crucial in determining the magnetic properties of the rare-earth manganites.\cite{Kimura1}

\begin{figure}
\includegraphics[width=2.5in]{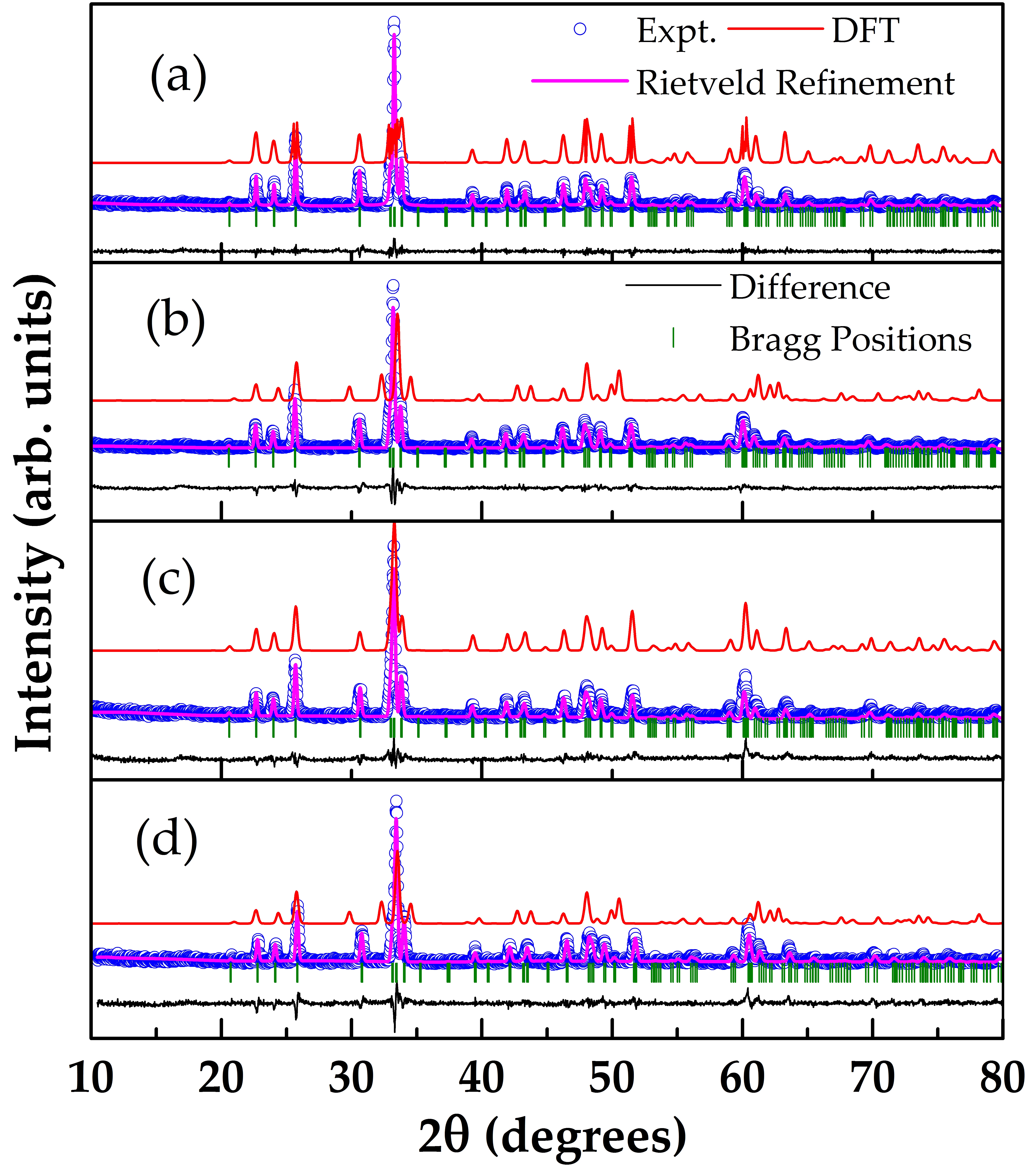}
\caption[Figure1] {(Color online) The simulated and measured XRD $\theta$--2$\theta$ scans of orthorhombic (a) TbMnO$_3$, (b) Tb$_{0.67}$Gd$_{0.33}$MnO$_3$, (c) Tb$_{0.67}$Dy$_{0.33}$MnO$_3$, and (d) Tb$_{0.67}$Ho$_{0.33}$MnO$_3$.

 {\label{fig:Figure1} } }
\end{figure}

\begin{figure}
\includegraphics[width=3.0in, height = 2in]{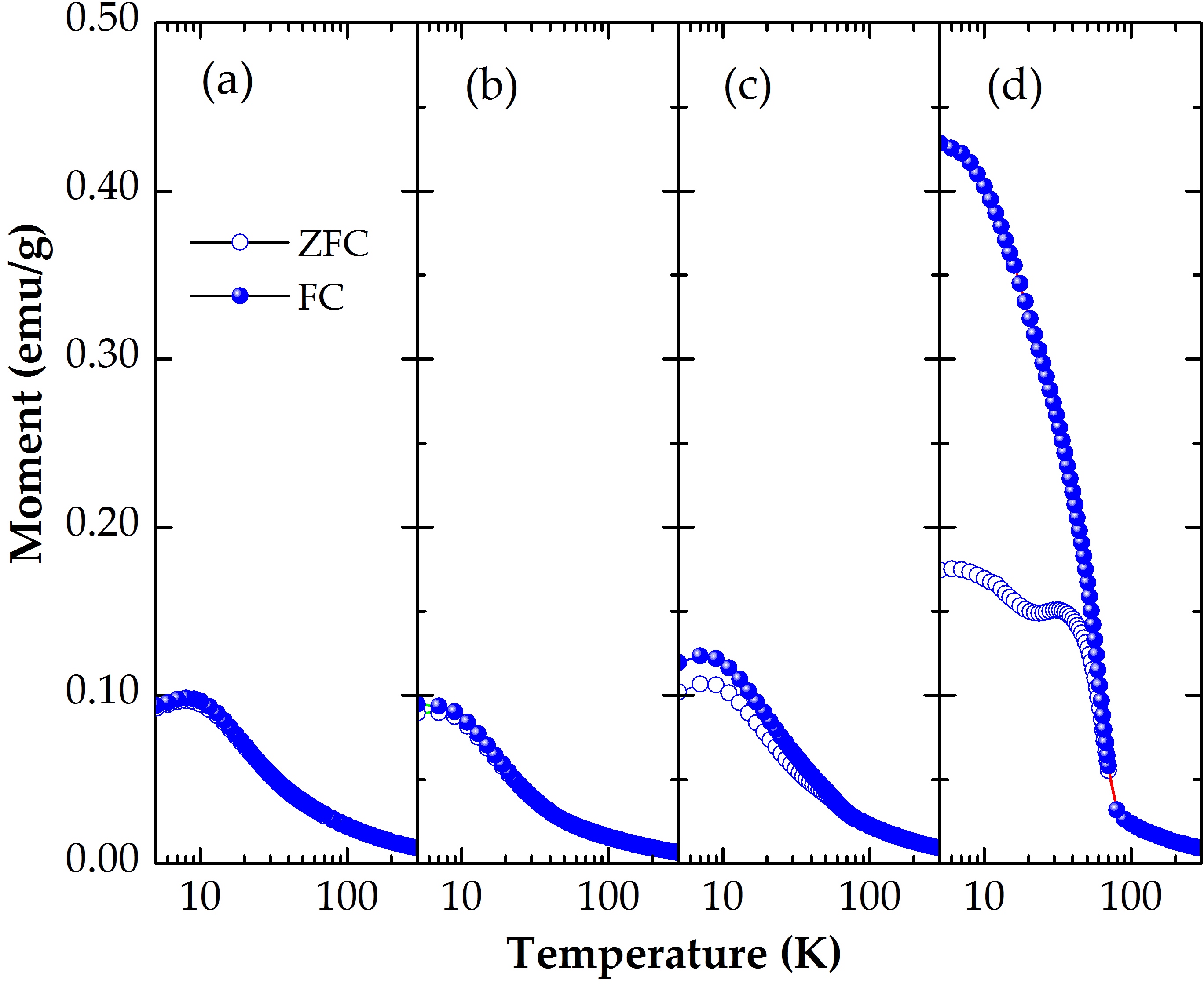}
\caption[Figure1] {(Color online) (\textbf{a}) The temperature dependent zero-field cooled (ZFC) and field cooled (FC) dc magnetization data for  (a) TbMnO$_3$, (b) Tb$_{0.67}$Gd$_{0.33}$MnO$_3$, (c) Tb$_{0.67}$Dy$_{0.33}$MnO$_3$, and (d) Tb$_{0.67}$Ho$_{0.33}$MnO$_3$ bulk samples.
 {\label{fig:Figure1} } }
\end{figure}

To validate our predicted results we measure the magnetic properties of these powder samples. The temperature dependent magnetization of the present TbMnO$_3$  and Tb$_{0.67}$\textit{A}$_{0.33}$MnO$_3$ (\textit{A} = Gd, Dy and Ho) samples is shown in Figure 5 in both zero-field-cooled (ZFC) and field-cooled (FC) conditions (with applied field of 50 Oe). The peak at $\sim$ 7 K in each of the present samples is due to the rare-earth magnetic ordering. No such features are observed for the paramagnetic to spiral antiferromagnetic and incommensurate to commensurate Mn magnetic transitions for pure and doped terbium manganites samples.\cite{Staruch-MCP}   A bifurcation of the ZFC and FC curves is observed at 60 K and 63 K for Tb$_{0.67}$Dy$_{0.33}$MnO$_3$ and Tb$_{0.67}$Ho$_{0.33}$MnO$_3$, respectively, suggesting a weak ferromagnetic component of the Mn ordering. The high temperature (paramagnetic) susceptibility data was fit to a Curie-Weiss law:
$\chi$=C/(T-${\theta}$),
where $\chi$ is the susceptibility, C is the Curie constant, and ${\theta}$ is the Weiss temperature. The effective magnetization are shown in Table 1. The negative Weiss temperatures observed in all of the present samples indicate antiferromagnetic ordering. The effective magnetic moments were calculated from the Curie constants. Gd substitution significantly reduces the effective magnetization, whereas Dy and Ho substitution only slightly reduce the effective magnetization as compared to pure TbMnO$_3$ . This is likely due to the high magnetic moments of Tb, Dy, and Ho compared to the moderate moment of Gd.\cite{Jensen} The measured magnetic moment is found to be in good agreement with our DFT calculated values (see Table 1). 

Above the Tb ordering temperature at 7 K, the Tb sub-lattice is expected to exhibit ordering as a consequence of J$_{Mn-Tb}$ exchange coupling.\cite{Lawes} Even below 7 K, the ordering of Tb is demonstrated to be related to the wave vector of Mn ordering in this temperature regime as a result of relatively strong Mn exchange field.\cite{Prokhnenko} Tb is thought to be a special case of intermediate coupling that leads to complex spin ordering behavior with respect to Mn sublattice, whereas Dy-Mn coupling is weak and Ho-Mn is very strong which leads to the wave vectors completely decoupled or coupled (respectively) down to zero temperature. Thus, we can expect that the substitution of 33\% Tb with another rare-earth ion will significantly impact the exchange coupling and consequently the ordering of the rare-earth moments at low temperatures. The modulation of the Mn-O-Mn bond angles and J$_{Mn-A}$ in the present samples may explain a significant difference in susceptibility and coercivity observed in the magnetic field dependent magnetization behavior (not shown), with stronger coupling leading to increased susceptibility at low applied fields. Gd substitution in TbMnO$_3$ slightly reduced the Mn-O-Mn bond angle and lead to slightly weakened magnetization values. Dy and Ho substitution both resulted in higher Mn-O-Mn bond angle than that in pure TbMnO$_3$ and enhanced  ferromagnetic component of the magnetization value. Additionally, the strong Ho-Mn interaction could account for the much larger increase in the magnetization value in case of Ho substitution, as compared to the weak Dy-Mn interaction present in the Dy substituted sample. However, any modification to the magnetic ordering of the Mn moments (\textit{i.e.} the weak ferromagnetism observed in Dy- and Ho-doped TbMnO$_3$ samples at higher temperatures) is expected to be a result of the theoretically and experimentally determined change in the Mn-O-Mn bond angles that will directly affect super-exchange interactions. 
 
In summary, we extend this dopant-mediated search for new possible multiferroic candidates by introducing rare-earth dopants in the parent material TbMnO$_3$. Using DFT based computations we predict that at low temperatures magnetic ground state of both pure and doped material systems is spiral-spin type of magnetic ordering. The predicted results are in favorable agreement with experimental measurements on the same material systems. Our theoretical and experimental research efforts have led to an improved understanding of both the spin structure and magnetization of doped systems. The major findings of this work are that the spiral AFM order observed in TbMnO$_3$ is stable with respect to the present A-site substitutions. These substitutions do, however, modify the Mn-O-Mn bond angles. The higher Mn-O-Mn bond angles, caused by the Dy and Ho substitutions, correspond with a stronger ferromagnetic component of the magnetic moment. This is a crucial finding that the magnetic order of TbMnO$_3$, which is known to cause the ferroelectricity, can be preserved while the net magnetic moment enhanced. This is an important step in the field of the magnetically driven ferroelectric class of magnetoelectric multiferroics, which have traditionally had low magnetic moments due to the predominantly antiferromagnetic order.\cite{Scott} In a broad sense, our research has led to an improved understanding of both dopant-medieated spin structure and magnetization in TbMnO$_3$.  However, further studies are needed to determine how these present substitutions could affect the ferroelectric polarization.

Partial computational support through a National Science Foundation Teragrid allocation is also gratefully acknowledged. Authors acknowledge the support of the National Science Foundation grant DMR-1310149 and UConn Large Grant.


\end{document}